\documentstyle[12pt]{article}

\title{Functional Tetrahedron Equation}

\author{
R.M. Kashaev\\
\small $\matrix{\noalign{\medskip}
\hbox{St.~Petersburg Branch of Steklov Mathematical Institute}\cr
\hbox{Fontanka 27, St.~Petersburg 191011, Russia.}\cr
\hbox{E-mail: {\tt kashaev@pdmi.ras.ru}}}$\\ \\
I.G. Korepanov\\
\small $\matrix{\noalign{\medskip}
\hbox{South Ural State University}\cr
\hbox{76 Lenin av., Chelyabinsk 454080, Russia}\cr
\hbox{E-mail: {\tt igor@prima.tu-chel.ac.ru}}}$\\ \\
S.M. Sergeev\\
\small $\matrix{\noalign{\medskip}
\hbox{Branch Institute for Nuclear Physics}\cr
\hbox{Protvino 142284, Russia.}\cr
\hbox{E-mail: {\tt sergeev\_ms@mx.ihep.su}}}$}

\date{January 1998}

\makeatletter

\def\be{\begin{equation}}
\def\ee{\end{equation}}

\def\Dcal{{\cal D}}

\long\def\@makecaption#1#2{\vskip 10\p@ \hbox to\hsize{\hfil#1\hfil}}

\makeatother

\begin{document}
\maketitle

\begin{abstract}
We describe a scheme of constructing classical integrable models in
$2+1$-dimen\-sional discrete space-time, based on the functional
tetrahedron equation---equation that makes manifest the symmetries of a
model in local form. We construct a very general ``block-matrix model''
together with its algebro-geometric solutions, study its various
particular cases, and also present a remarkably simple scheme of quantization
for one of those cases.
\end{abstract}

\section*{Introduction}

Consider two electric devices, each made of three resistors and with three
outer contacts, and whose diagrams are triangle and star, respectively.
It is well known that they are equivalent iff the following relations for
resistances hold:
\be
r_jr'_j=r'_1r'_2+r'_2r'_3+r'_3r'_1={r_1r_2r_3\over r_1+r_2+r_3},
\quad j=1,2,3,
\label{ieq s-t electric}
\ee
where resistances $r_j$ apply to the triangle, and resistances
$r'_j$---to the star.
These relations enable us to find $r'_j$ for any given triple of~$r_j$,
and vice versa.

We will represent such electric devices diagrammatically in a simplified
manner (in comparison with usual electric diagrams): resistors will be
represented as edges of a graph, outer contacts---as blank circles, and
inner contacts---as filled circles. In this way, the
transformation~(\ref{ieq s-t electric}) is represented in
Figure~\ref{iris1-2}.
\begin{figure}[htp]
\begin{center}
\unitlength=0.50mm
\special{em:linewidth 0.4pt}
\linethickness{0.4pt}
\begin{picture}(162.00,57.00)
\put(4.00,3.00){\line(3,5){30.00}}
\put(36.00,53.00){\line(3,-5){30.00}}
\put(66.00,2.00){\line(-1,0){62.00}}
\put(96.00,3.00){\line(5,3){30.00}}
\put(128.00,21.00){\line(5,-3){30.00}}
\put(127.00,23.00){\line(0,1){30.00}}
\put(20.00,32.00){\makebox(0,0)[rb]{$r_3$}}
\put(50.00,32.00){\makebox(0,0)[lb]{$r_1$}}
\put(35.00,5.00){\makebox(0,0)[cb]{$r_2$}}
\put(128.00,39.00){\makebox(0,0)[lc]{$r'_2$}}
\put(106.00,12.00){\makebox(0,0)[rb]{$r'_1$}}
\put(145.00,12.00){\makebox(0,0)[lb]{$r'_3$}}
\put(81.00,29.00){\vector(-1,0){9.00}}
\put(79.00,29.00){\vector(1,0){9.00}}
\put(2.00,2.00){\circle{4.00}}
\put(35.00,55.00){\circle{4.00}}
\put(68.00,2.00){\circle{4.00}}
\put(127.00,22.00){\circle*{4.00}}
\put(160.00,2.00){\circle{4.00}}
\put(94.00,2.00){\circle{4.00}}
\put(127.00,55.00){\circle{4.00}}
\end{picture}
\end{center}
\caption{}
\label{iris1-2}
\end{figure}

Note that formulae~(\ref{ieq s-t electric}) admit the interchange
$$
r_j \leftrightarrow {1\over r'_j}.
$$
This leads to the fact that the ``star--triangle'' and ``triangle--star''
transformations, after a simple change of variables, can be expressed
as the same transformation. The exact formulation of this will be given
below in subsection~\ref{subsec epsilon}.

Consider now the diagrams in Figure~\ref{iris3}.
\begin{figure}[htp]
\begin{center}
\unitlength=1mm
\special{em:linewidth 0.4pt}
\linethickness{0.4pt}
\begin{picture}(91.00,60.00)
\put(1.00,35.00){\circle{2.00}}
\put(2.00,35.00){\line(1,0){13.00}}
\put(16.00,35.00){\circle*{2.00}}
\put(16.70,34.30){\line(1,-1){8.60}}
\put(26.00,25.00){\circle*{2.00}}
\put(26.00,11.00){\circle{2.00}}
\put(26.00,12.00){\line(0,1){12.00}}
\put(36.00,35.00){\circle{2.00}}
\put(26.00,45.00){\circle{2.00}}
\put(16.70,35.70){\line(1,1){8.60}}
\put(26.70,44.30){\line(1,-1){8.60}}
\put(35.30,34.30){\line(-1,-1){8.60}}
\put(56.00,35.00){\circle{2.00}}
\put(66.00,25.00){\circle{2.00}}
\put(76.00,35.00){\circle*{2.00}}
\put(66.00,45.00){\circle*{2.00}}
\put(66.00,46.00){\line(0,1){12.00}}
\put(66.00,59.00){\circle{2.00}}
\put(77.00,35.00){\line(1,0){12.00}}
\put(90.00,35.00){\circle{2.00}}
\put(75.30,35.70){\line(-1,1){8.60}}
\put(65.30,44.30){\line(-1,-1){8.60}}
\put(56.70,34.30){\line(1,-1){8.60}}
\put(66.70,25.70){\line(1,1){8.60}}
\put(1.00,37.00){\makebox(0,0)[lb]{$1$}}
\put(26.00,47.00){\makebox(0,0)[cb]{$2$}}
\put(36.00,37.00){\makebox(0,0)[lb]{$3$}}
\put(28.00,11.00){\makebox(0,0)[lb]{$4$}}
\put(26.00,2.00){\makebox(0,0)[cb]{(a)}}
\put(66.00,2.00){\makebox(0,0)[cb]{(b)}}
\put(55.00,37.00){\makebox(0,0)[rb]{$1$}}
\put(68.00,59.00){\makebox(0,0)[lt]{$2$}}
\put(90.00,37.00){\makebox(0,0)[rb]{$3$}}
\put(68.00,23.00){\makebox(0,0)[lt]{$4$}}
\end{picture}
\end{center}
\caption{}
\label{iris3}
\end{figure}
We can transform diagram (a) into diagram (b) in four steps, each being
a star--triangle or triangle--star transformation for some three resistors.
What is important, we can do it in {\em two ways\/}, represented in
Figure~\ref{iris4}
\begin{figure}[htp]
\begin{center}
\unitlength=1mm
\special{em:linewidth 0.4pt}
\linethickness{0.4pt}
\begin{picture}(106.00,56.00)
\put(12.00,0.00){\line(0,1){7.00}}
\put(12.00,7.00){\line(1,1){5.00}}
\put(17.00,12.00){\line(-1,1){5.00}}
\put(12.00,17.00){\line(-1,-1){5.00}}
\put(7.00,12.00){\line(1,-1){5.00}}
\put(7.00,12.00){\line(-1,0){7.00}}
\put(19.00,12.00){\vector(1,0){3.00}}
\put(24.00,12.00){\line(1,0){17.00}}
\put(41.00,12.00){\line(-1,1){5.00}}
\put(36.00,17.00){\line(-1,-1){5.00}}
\put(31.00,12.00){\line(1,-1){5.00}}
\put(36.00,7.00){\line(1,1){5.00}}
\put(43.00,12.00){\vector(1,0){3.00}}
\put(48.00,12.00){\line(1,0){7.00}}
\put(55.00,12.00){\line(1,-1){5.00}}
\put(60.00,7.00){\line(1,1){5.00}}
\put(65.00,12.00){\line(-1,1){5.00}}
\put(60.00,17.00){\line(-1,-1){5.00}}
\put(60.00,17.00){\line(0,1){7.00}}
\put(67.00,12.00){\vector(1,0){3.00}}
\put(72.00,12.00){\line(1,1){5.00}}
\put(77.00,17.00){\line(1,-1){5.00}}
\put(82.00,12.00){\line(-1,-1){5.00}}
\put(77.00,7.00){\line(-1,1){5.00}}
\put(77.00,7.00){\line(0,1){17.00}}
\put(84.00,12.00){\vector(1,0){3.00}}
\put(89.00,12.00){\line(1,-1){5.00}}
\put(94.00,7.00){\line(1,1){5.00}}
\put(99.00,12.00){\line(-1,1){5.00}}
\put(94.00,17.00){\line(-1,-1){5.00}}
\put(99.00,12.00){\line(1,0){7.00}}
\put(94.00,17.00){\line(0,1){7.00}}
\put(12.00,32.00){\line(0,1){7.00}}
\put(12.00,39.00){\line(1,1){5.00}}
\put(17.00,44.00){\line(-1,1){5.00}}
\put(12.00,49.00){\line(-1,-1){5.00}}
\put(7.00,44.00){\line(1,-1){5.00}}
\put(7.00,44.00){\line(-1,0){7.00}}
\put(19.00,44.00){\vector(1,0){3.00}}
\put(84.00,44.00){\vector(1,0){3.00}}
\put(89.00,44.00){\line(1,-1){5.00}}
\put(94.00,39.00){\line(1,1){5.00}}
\put(99.00,44.00){\line(-1,1){5.00}}
\put(94.00,49.00){\line(-1,-1){5.00}}
\put(99.00,44.00){\line(1,0){7.00}}
\put(94.00,49.00){\line(0,1){7.00}}
\put(24.00,44.00){\line(1,-1){5.00}}
\put(29.00,39.00){\line(1,1){5.00}}
\put(34.00,44.00){\line(-1,1){5.00}}
\put(29.00,49.00){\line(-1,-1){5.00}}
\put(29.00,49.00){\line(0,-1){17.00}}
\put(36.00,44.00){\vector(1,0){3.00}}
\put(41.00,44.00){\line(1,-1){5.00}}
\put(46.00,39.00){\line(1,1){5.00}}
\put(51.00,44.00){\line(-1,1){5.00}}
\put(46.00,49.00){\line(-1,-1){5.00}}
\put(46.00,39.00){\line(0,-1){7.00}}
\put(51.00,44.00){\line(1,0){7.00}}
\put(60.00,44.00){\vector(1,0){3.00}}
\put(65.00,44.00){\line(1,1){5.00}}
\put(70.00,49.00){\line(1,-1){5.00}}
\put(75.00,44.00){\line(-1,-1){5.00}}
\put(70.00,39.00){\line(-1,1){5.00}}
\put(65.00,44.00){\line(1,0){17.00}}
\end{picture}
\end{center}
\caption{}
\label{iris4}
\end{figure}
(where the obvious arrangement of blank and filled circles is not shown).

It can be seen from ``electrical'' argument that the two ways in
Figure~\ref{iris4} must lead to the same result. Mathematically, this means
that the transformation~(\ref{ieq s-t electric}) is very specific:
it satisfies (when written as in subsection~\ref{subsec epsilon} below)
the {\em functional tetrahedron equation\/} (FTE).

In this paper, we will try to show that FTE constitues the basis of
$2+1$-dimensional integrability. We will explain how to generalize
the simple ``electrical'' construction for its solutions in such way that
it will give not only a huge amount of new integrable classical systems,
but even integrable quantum systems. The contents of the rest of this paper
is as follows. In section~\ref{sec straight-string} we consider the most
general abstract model where the FTE naturally arises.
In section~\ref{sec block-matrix} we present a concrete, but still very
general incarnation of the abstract model, together with the
algebro-geometric method of constructing its solutions.
In section~\ref{sec one-parametric} we
give a list of solutions representing the simplest ``one-parametric
reductions'' of the general model. In section~\ref{sec miwa} we show
that the ``electric'' model described above is connected with the
well-known bilinear Miwa equation. In section~\ref{sec quant}
we present an amazingly direct quantization for one of our
``one-parametric'' models, using, essentially, the same FTE.
Finally, we discuss our results in section~\ref{discussion}.

\section{Classical straight-string model and functional tetrahedron
equation}
\label{sec straight-string}

Let there be several straight strings, i.e.\ oriented straight lines
(with the ``positive direction'' indicated in each line),
in a plane. Let there be among those strings ({\it i\/})\,no
coinciding ones, ({\it ii\/})\,no three of them intersecting in one
point, and ({\it iii\/})\,let the orientations of strings be
{\em consistent\/} in the following sense: moving from any point
along the strings in positive directions,
one cannot return in the same point.
Let us attribute to each point of intersection of two strings,
e.g.\ strings $a$ and $b$, some object $X_{ab}$. Now let us allow
each string to move, remaining parallel to its initial position,
and pass through points of intersection of other strings, but so that
({\it i\/})\,two parallel strings never go through each other and
({\it ii\/})\,at no moment there should
be four strings intersecting in one point.

Let us assume that an object $X_{ab}$ changes only if some string $c$
goes through the intersection point of $a$ and $b$. To be exact, in
such case the {\em triple\/} $(X_{ab},X_{ac},X_{bc})$ is transformed
into a new triple $(X'_{ab},X'_{ac},X'_{bc})$ according to some fixed
rule. When we consider such a transformation, we think of
both those triples as ordered in the following way:
one can get from $X_{ab}$ to $X_{ac}$ and then to $X_{bc}$
moving along the strings in positive directions, while
 from $X'_{ab}$ to $X'_{ac}$ and then to $X'_{bc}$---in negative directions.
This can be represented graphically as turning the triangle inside
out, see Figure~\ref{ifig1}.
\begin{figure}[htp]
\begin{center}
\unitlength=1.00mm
\special{em:linewidth 0.4pt}
\linethickness{0.4pt}
\begin{picture}(125.00,63.00)
\put(0.00,35.00){\line(2,-1){50.00}}
\put(5.00,15.00){\line(3,2){54.00}}
\put(35.00,0.00){\line(1,3){21.00}}
\put(80.00,0.00){\line(1,3){21.00}}
\put(70.00,5.00){\line(3,2){54.00}}
\put(75.00,55.00){\line(2,-1){50.00}}
\put(95.00,45.00){\circle*{1.00}}
\put(115.00,35.00){\circle*{1.00}}
\put(85.00,15.00){\circle*{1.00}}
\put(40.00,15.00){\circle*{1.00}}
\put(50.00,45.00){\circle*{1.00}}
\put(20.00,25.00){\circle*{1.00}}
\put(4.00,33.00){\vector(2,-1){2.00}}
\put(36.00,3.00){\vector(1,3){1.00}}
\put(81.00,52.00){\vector(2,-1){2.00}}
\put(81.00,3.00){\vector(1,3){1.00}}
\put(3.00,35.00){\makebox(0,0)[cb]{$a$}}
\put(7.00,18.00){\makebox(0,0)[rb]{$b$}}
\put(34.00,4.00){\makebox(0,0)[rc]{$c$}}
\put(20.00,28.00){\makebox(0,0)[cb]{$X_{ab}$}}
\put(44.00,17.00){\makebox(0,0)[lc]{$X_{ac}$}}
\put(51.00,43.00){\makebox(0,0)[lt]{$X_{bc}$}}
\put(79.00,55.00){\makebox(0,0)[cb]{$a$}}
\put(73.00,9.00){\makebox(0,0)[rb]{$b$}}
\put(79.00,2.00){\makebox(0,0)[rb]{$c$}}
\put(91.00,43.00){\makebox(0,0)[rc]{$X'_{ac}$}}
\put(114.00,38.00){\makebox(0,0)[cb]{$X'_{ab}$}}
\put(87.00,13.00){\makebox(0,0)[lt]{$X'_{bc}$}}
\put(60.00,29.00){\vector(1,0){9.00}}
\put(8.00,17.00){\vector(3,2){3.00}}
\put(73.00,7.00){\vector(3,2){3.00}}
\end{picture}
\end{center}
\caption{}
\label{ifig1}
\end{figure}

We will say that the {\em operator} $R$ acts on the triples of objects:
\be
R:\quad (X_{ab},X_{ac},X_{bc}) \to (X'_{ab},X'_{ac},X'_{bc}).
\label{ieq R}
\ee
Note that in Figure~\ref{ifig1} the point of intersection of strings
$a$ and~$b$ goes from the left part of the plane, with respect to
orientation of~$c$, to the right part. If we make in Figure~\ref{ifig1}
the changes $c\leftrightarrow a$, $X_{\ldots}\leftrightarrow X'_{\ldots}$,
${\rm LHS}\leftrightarrow {\rm RHS}$, we will see that for the case
when the point of intersection of strings
$a$ and~$b$ goes from the right part of the plane, with respect to
orientation of~$c$, to the left part, $R$ should be replaced by
$P_{13}R^{-1}P_{13}$, where $P_{13}$ stands for the interchange of the
first and the third objects in a triple. In other respects,
we assume $R$ be the same for all triples of strings,
i.e.\ not depending, say, on the angles between them or whatever.
At the same time, we will not exclude the situation where $R$ is not
single-valued---in such case we will say that the ``primed triple''
is determined up to a ``gauge equivalence''.

If there are many enough non-parallel strings in our model, there exist
many possibilities for their movement. It is natural to regard as
``integrable'' such a model where to a passage from one string
configuration to another there will correspond a transformation
of objects $X_{ab}$ not depending of the details of that passage,
e.g.\ which string was the first to pass through the intersection point
of two others and so on. For this to hold, it is enough to require
that the operation of turning a triangle inside out, as in
Figure~\ref{ifig1}, commute with passing a fourth string across
the whole triangle. In other words, the two ways of transforming
the left side of Figure~\ref{ifig2}
\begin{figure}[htp]
\begin{center}
\unitlength=1.00mm
\special{em:linewidth 0.4pt}
\linethickness{0.4pt}
\begin{picture}(115.00,53.00)
\put(45.00,0.00){\line(0,1){53.00}}
\put(1.00,8.00){\line(2,1){50.00}}
\put(0.00,15.00){\line(1,0){51.00}}
\put(23.00,1.00){\line(1,2){26.00}}
\put(70.00,0.00){\line(0,1){53.00}}
\put(66.00,0.00){\line(1,2){26.00}}
\put(64.00,38.00){\line(1,0){51.00}}
\put(64.00,20.00){\line(2,1){50.00}}
\put(4.00,15.00){\vector(1,0){1.00}}
\put(3.00,9.00){\vector(2,1){2.00}}
\put(24.00,3.00){\vector(1,2){1.00}}
\put(45.00,3.00){\vector(0,1){1.00}}
\put(67.00,2.00){\vector(1,2){1.00}}
\put(70.00,2.00){\vector(0,1){1.00}}
\put(66.00,21.00){\vector(2,1){2.00}}
\put(66.00,38.00){\vector(1,0){1.00}}
\put(3.00,16.00){\makebox(0,0)[cb]{$a$}}
\put(3.00,10.00){\makebox(0,0)[rb]{$b$}}
\put(23.00,4.00){\makebox(0,0)[rb]{$c$}}
\put(43.00,3.00){\makebox(0,0)[rc]{$d$}}
\put(66.00,40.00){\makebox(0,0)[cb]{$a$}}
\put(67.00,23.00){\makebox(0,0)[rb]{$b$}}
\put(67.00,6.00){\makebox(0,0)[rb]{$c$}}
\put(72.00,4.00){\makebox(0,0)[lc]{$d$}}
\put(15.00,15.00){\circle*{1.00}}
\put(30.00,15.00){\circle*{1.00}}
\put(45.00,15.00){\circle*{1.00}}
\put(45.00,30.00){\circle*{1.00}}
\put(45.00,45.00){\circle*{1.00}}
\put(35.00,25.00){\circle*{1.00}}
\put(70.00,23.00){\circle*{1.00}}
\put(70.00,38.00){\circle*{1.00}}
\put(85.00,38.00){\circle*{1.00}}
\put(100.00,38.00){\circle*{1.00}}
\put(80.00,28.00){\circle*{1.00}}
\put(70.00,8.00){\circle*{1.00}}
\put(15.00,17.00){\makebox(0,0)[cb]{$1$}}
\put(34.00,26.00){\makebox(0,0)[rb]{$3$}}
\put(43.00,45.00){\makebox(0,0)[rc]{$6$}}
\put(46.00,29.00){\makebox(0,0)[lt]{$5$}}
\put(46.00,14.00){\makebox(0,0)[lt]{$4$}}
\put(31.00,13.00){\makebox(0,0)[lt]{$2$}}
\put(71.00,39.00){\makebox(0,0)[lb]{$4$}}
\put(84.00,39.00){\makebox(0,0)[rb]{$2$}}
\put(99.00,41.00){\makebox(0,0)[rc]{$1$}}
\put(81.00,27.00){\makebox(0,0)[lt]{$3$}}
\put(72.00,11.00){\makebox(0,0)[lt]{$6$}}
\put(53.00,26.00){\vector(1,0){9.00}}
\put(71.00,22.00){\makebox(0,0)[lt]{$5$}}
\end{picture}
\end{center}
\caption{}
\label{ifig2}
\end{figure}
into its right side, one of which
starts with turning inside out triangle $356$ and the other
one---triangle $123$, must lead to equal results. This condition
is the {\em functional tetrahedron equation\/} (FTE) to which this
paper is devoted:
\be
R_{123} \circ R_{145} \circ R_{246} \circ R_{356} =
R_{356} \circ R_{246} \circ R_{145} \circ R_{123}.
\label{ieq1}
\ee
Here, of course, $R_{123}$ acts only on the triple
$$
(X_1,X_2,X_3) \stackrel{\rm def}{=} (X_{ab},X_{ac},X_{bc})
$$
and so on.

The diagrammatic representation of equation (\ref{ieq1}) itself
is given in Figure~\ref{ifig3}.
\begin{figure}[htp]
\begin{center}
\unitlength=1.00mm
\special{em:linewidth 0.4pt}
\linethickness{0.4pt}
\begin{picture}(93.00,55.00)
\put(0.00,35.00){\line(1,0){40.00}}
\put(20.00,5.00){\line(0,1){50.00}}
\put(0.00,30.00){\line(2,1){36.00}}
\put(4.00,48.00){\line(2,-1){36.00}}
\put(33.00,41.00){\line(-1,-2){17.00}}
\put(24.00,7.00){\line(-1,2){17.00}}
\put(73.00,5.00){\line(0,1){50.00}}
\put(53.00,25.00){\line(1,0){40.00}}
\put(60.00,19.00){\line(1,2){17.00}}
\put(69.00,53.00){\line(1,-2){17.00}}
\put(93.00,30.00){\line(-2,-1){36.00}}
\put(89.00,12.00){\line(-2,1){36.00}}
\put(6.00,47.00){\vector(2,-1){2.00}}
\put(2.00,35.00){\vector(1,0){1.00}}
\put(2.00,31.00){\vector(2,1){2.00}}
\put(17.00,9.00){\vector(1,2){1.00}}
\put(20.00,7.00){\vector(0,1){1.00}}
\put(23.00,9.00){\vector(-1,2){1.00}}
\put(55.00,29.00){\vector(2,-1){2.00}}
\put(55.00,25.00){\vector(1,0){1.00}}
\put(61.00,21.00){\vector(1,2){1.00}}
\put(59.00,13.00){\vector(2,1){2.00}}
\put(73.00,7.00){\vector(0,1){1.00}}
\put(85.00,21.00){\vector(-1,2){1.00}}
\put(6.00,48.00){\makebox(0,0)[lb]{$X_1$}}
\put(2.00,36.00){\makebox(0,0)[cb]{$X_2$}}
\put(0.00,31.00){\makebox(0,0)[rb]{$X_4$}}
\put(15.00,8.00){\makebox(0,0)[rb]{$X_3$}}
\put(20.00,5.00){\makebox(0,0)[lt]{$X_5$}}
\put(24.00,9.00){\makebox(0,0)[lb]{$X_6$}}
\put(23.00,15.00){\makebox(0,0)[lc]{$R_{356}$}}
\put(30.00,30.00){\makebox(0,0)[lt]{$R_{123}$}}
\put(21.00,46.00){\makebox(0,0)[lb]{$R_{145}$}}
\put(12.00,30.00){\makebox(0,0)[rt]{$R_{246}$}}
\put(55.00,24.00){\makebox(0,0)[ct]{$X_2$}}
\put(60.00,19.00){\makebox(0,0)[rt]{$X_3$}}
\put(57.00,11.00){\makebox(0,0)[ct]{$X_4$}}
\put(72.00,5.00){\makebox(0,0)[rc]{$X_5$}}
\put(87.00,19.00){\makebox(0,0)[lc]{$X_6$}}
\put(65.00,30.00){\makebox(0,0)[rb]{$R_{123}$}}
\put(74.00,15.00){\makebox(0,0)[lt]{$R_{145}$}}
\put(81.00,31.00){\makebox(0,0)[lb]{$R_{246}$}}
\put(75.00,45.00){\makebox(0,0)[lc]{$R_{356}$}}
\put(46.00,26.00){\makebox(0,0)[cc]{$=$}}
\put(53.00,31.00){\makebox(0,0)[rt]{$X_1$}}
\put(10.00,35.00){\circle*{1.30}}
\put(20.00,40.00){\circle*{1.30}}
\put(30.00,35.00){\circle*{1.30}}
\put(20.00,15.00){\circle*{1.30}}
\put(63.00,25.00){\circle*{1.30}}
\put(73.00,20.00){\circle*{1.30}}
\put(83.00,25.00){\circle*{1.30}}
\put(73.00,45.00){\circle*{1.30}}
\end{picture}
\end{center}
\caption{}
\label{ifig3}
\end{figure}

To conclude this section, note that we could assosiate $X$-like objects
not to the intersection points of strings, but to the segments in which
the strings divide each other, or to the domains of the plane
in which it is divided by the strings. It would lead us to other
versions of the functional tetrahedron equation, in complete analogy with
the quantum case~\cite{B-S--J-M,hietarinta-labeling}.
It is natural to call equation~(\ref{ieq1}) the {\em vertex type\/}
FTE. In this paper we will be dealing only with this type of equations.

\section{The block-matrix model and its algebro-geometric solutions}
\label{sec block-matrix}

\subsection{Formulation of the model}
\label{subsec formulation}

Let now to each of the strings correspond a finite-dimensional complex
linear space, e.g.\ to the string~$a$---the space~$V_a$.
We will also associate with a set of strings the {\em direct sum\/}
of their linear spaces (and not the tensor product, in contrast with
the usual theory of Yang--Baxter equation). The objects $X_{ab}$
will be {\em linear operators\/} in $V_a\oplus V_b$. Fixing once
and for ever the bases in all $V_a$, we will identify those operators
with block matrices of the type
$$
X=\pmatrix{A&B\cr C&D},
$$
where $A$ acts within $V_a$, $B$---from $V_b$ to $V_a$, etc.
The ``primed'' operators---the result of the action of
operator $R$~(\ref{ieq R})---will be determined from the equation
\be
X_{ab}X_{ac}X_{bc}=X'_{bc}X'_{ac}X'_{ab},
\label{ieq2}
\ee
which in expanded form is written as
\begin{eqnarray}
&&
\pmatrix{ A_1&B_1&{\bf 0}\cr C_1&D_1&{\bf 0} \cr 
{\bf 0}&{\bf 0}&{\bf 1} }
\pmatrix{ A_2&{\bf 0}&B_2\cr {\bf 0}&{\bf 1}&{\bf 0}\cr 
C_2&{\bf 0}&D_2 }
\pmatrix{ {\bf 1}&{\bf 0}&{\bf 0}\cr 
{\bf 0}&A_3&B_3\cr {\bf 0}&C_3&D_3 }
\nonumber\\[0.5\normalbaselineskip]
&& =
\pmatrix{ {\bf 1}&{\bf 0}&{\bf 0}\cr 
{\bf 0}&A'_3&B'_3\cr {\bf 0}&C'_3&D'_3 }
\pmatrix{ A'_2&{\bf 0}&B'_2\cr {\bf 0}&{\bf 1}&{\bf 0}\cr 
C'_2&{\bf 0}&D'_2 }
\pmatrix{ A'_1&B'_1&{\bf 0}\cr C'_1&D'_1&{\bf 0} \cr 
{\bf 0}&{\bf 0}&{\bf 1} },
\label{ieq pererazlozhenie}
\end{eqnarray}
i.e.\ we identify each of the operators
$$
X_{ab}=X_1=\pmatrix{A_1&B_1\cr C_1&D_1},\quad\ldots\quad,\quad
X'_{bc}=X'_3=\pmatrix{A'_3&B'_3\cr C'_3&D'_3}
$$
with the direct sum of itself and the unity operator in the lacking
space.

\medskip
{\bf Remark. }If we now want to consider Figure~\ref{ifig1} as the
illustration to formula~(\ref{ieq2}), we have to take into account
that the operators in~(\ref{ieq2}) act {\em against\/} the arrows
in Figure~\ref{ifig1}. On the other hand, in Figure~\ref{ifig3}
the arrows point in the {\em same\/} direction where the (nonlinear)
operators~$R$ act.
\medskip

In case if the operators in the LHS of equation (\ref{ieq2}) are
generic enough, that equation is solvable with respect to the ``primed''
operators, and those latter are determined uniquely up to the following
obvious block-diagonal ``gauge transformations'':
\begin{eqnarray*}
\pmatrix{A'_3&B'_3\cr C'_3&D'_3}&\to&
\pmatrix{A'_3&B'_3\cr C'_3&D'_3}\pmatrix{L&\bf 0\cr \bf 0&M},\\
\pmatrix{A'_2&B'_2\cr C'_2&D'_2}&\to&
\pmatrix{\bf 1&\bf 0\cr \bf 0&M^{-1}}\pmatrix{A'_2&B'_2\cr C'_2&D'_2}
\pmatrix{K&\bf 0\cr \bf 0&\bf 1},\\
\pmatrix{A'_1&B'_1\cr C'_1&D'_1}&\to&
\pmatrix{K^{-1}&\bf 0\cr \bf 0&L^{-1}}\pmatrix{A'_1&B'_1\cr C'_1&D'_1}.
\end{eqnarray*}
The details can be found in \cite{korepanov dis}, chapter~2.

As for the functional tetrahedron equation, it follows in our
block-matrix model from the fact that the product of {\em six\/}
$X$-matrices corresponding to the RHS of Figure~\ref{ifig2}
determines those matrices as well (again up to obvious block-diagonal gauge
transformations). The proof of this statement is a routine exercise
in matrix algebra (to begin, one can unite, for a while, the
strings $c$ and $d$ in the RHS of Figure~\ref{ifig2} in one ``thick''
string, and then apply several times the uniqueness, up to gauge
transformations, of factorizing the product of three matrices).

\subsection{Algebro-geometric solutions}

Let us take some (smooth irreducible) algebraic curve~$\Gamma$
of genus~$g$. Let us attach to each string~$a$ two effective divisors
(i.e.\ finite sets of points of~$\Gamma$
with {\em positive\/} multiplicities)
$\Dcal'_a$ and $\Dcal''_a$ of the same degree, equal to $\dim V_a$
(the dimension of linear space associated with~$a$).
For two non-parallel strings $a$ and $b$, we will
require that no point enter at the same time in both $\Dcal'_a$
and $\Dcal''_b$ (as well as in both $\Dcal''_a$ and $\Dcal'_b$).
The meaning of this will be clear later.

Next, let us
put in correspondence to each point~$A$ of the plane where the strings live
a divisor $\Dcal^A$ according to the following rules. First,
for each string~$a$, if $A$ lies to the left of $a$ (with respect to its
orientation), take divisor $\Dcal'_a$, if $A$ lies to the
right of $a$, take divisor $\Dcal''_a$, if $A$ lies in $a$, take the zero
divisor, and denote that as~$\Dcal_a^A$. Then set
$\Dcal^A=\sum_{{\rm all\hphantom{i}} a} \Dcal_a^A$.

We will only be interested in divisors corresponding this way to (inner
points of) string
segments limited by string intersection points, and to those
intersection points themselves. For example, if there are only two
strings, to their halves and to the intersection point will correspond
divisors as in Figure~\ref{ifig4}.
\begin{figure}[htp]
\begin{center}
\unitlength=1.00mm
\special{em:linewidth 0.4pt}
\linethickness{0.4pt}
\begin{picture}(65.00,55.00)
\put(10.00,5.00){\line(1,1){50.00}}
\put(5.00,30.00){\line(1,0){60.00}}
\put(55.00,50.00){\vector(1,1){1.00}}
\put(59.00,30.00){\vector(1,0){1.00}}
\put(4.00,30.00){\makebox(0,0)[rc]{$a$}}
\put(9.00,4.00){\makebox(0,0)[rt]{$b$}}
\put(15.00,32.00){\makebox(0,0)[cb]{${\cal D}'_b$}}
\put(53.00,32.00){\makebox(0,0)[cb]{${\cal D}''_b$}}
\put(48.00,45.00){\makebox(0,0)[rb]{${\cal D}'_a$}}
\put(19.00,16.00){\makebox(0,0)[rb]{${\cal D}''_a$}}
\put(34.00,28.00){\makebox(0,0)[lt]{${\cal D}_{\rm
in\hphantom{i}the\hphantom{i}vertex}=0$}}
\put(35.00,30.00){\circle*{1.00}}
\end{picture}
\end{center}
\caption{}
\label{ifig4}
\end{figure}

Take now an arbitrary (generic) divisor $\Dcal$ of degree
$$
\deg \Dcal=\sum_{{\rm over\;all\;}a} \dim V_a +g-1.
$$
If to a given segment $l$ of a string~$a$ corresponds a divisor~$\Dcal_l$,
let us also put in correspondence to that segment the linear space
of meromorphic functions~$f$ on~$\Gamma$ whose divisor $(f)$ (the
difference between the zero divisor and the pole divisor) obeys
the condition
$$
(f)+\Dcal-\Dcal_l \ge 0
$$
(this means that the singularities of $f$ are controlled by
divisor~$\Dcal$ and, moreover, $f$ must have zeroes in the points
of divisor~$\Dcal_l$).
It is easy to see that the space of such functions is isomorphic to
(i.e.\ has the same dimension as) $V_a$, so we will identify it with $V_a$,
using for that an arbitrary isomorphism.

We will need the following explicit description of isomorphisms of this
kind. The space~$V_a$ itself, whose basis, as we have agreed in the beginning
of subsection~\ref{subsec formulation}, is fixed, can be imagined as
comprised of
column vectors (that we will write sometimes as transposed row vector,
using the symbol~$\rm T$ of matrix transposing) of height $n_a=\dim V_a$.
Whenever $V_a$ should be identified with some space of meromorphic functions,
we should choose in the latter a basis $(f_1,\ldots,f_{n_a})$, and after
that put in correspondence to a column vector
$(\alpha_1,\ldots,\alpha_{n_a})^{\rm T}\in V_a$ the function
$\alpha_1 f_1+\ldots+\alpha_{n_a}f_{n_a}$.

Thus, in the different segments of string~$a$, the space~$V_a$
is identified with {\em different\/} spaces of meromorphic functions
by arbitrary but fixed isomorphisms. Every time as, following the
movement of the strings, the limiting points of a segment pass through
each other, the isomorphism for the ``turned inside out'' segment
is chosen, of course, anew.

Let strings $a$ and $b$ intersect in a vertex~$A$. Let us denote $\Dcal_A$
the divisor corresponding to~$A$ according to the rules described above.
The vertex~$A$ has two incoming edges (string segments {\em preceding\/}
$A$, according to the given positive directions in the strings)
and two outgoing
ones. It is easily seen that the sum of linear spaces of meromorphic
functions corresponding to the incoming edges consists of such
functions~$f$ that
$$
(f)+\Dcal-\Dcal_A\ge 0
$$
and is {\em direct\/} (the particular case shown in Figure~\ref{ifig4}
can be useful to quickly understand this. Recall the requirement
that no point enter in both $\Dcal''_a$ and $\Dcal'_b$).
Exactly the same applies to the outgoing edges, hence the sums
of ``incoming'' and ``outgoing'' spaces coincide.

The operator $X_{ab}$ is now constructed as follows. Take some vector
$\Phi \in V_a \oplus V_b$ and represent it as a sum
$$
\Phi=\Phi_a+\Phi_b,\qquad \Phi_a\in V_a,\quad \Phi_b\in V_b.
$$
Identify vectors $\Phi_a$ and $\Phi_b$ with meromorphic functions
$\varphi_a$ and $\varphi_b$ using the discussed above isomorphisms for the
{\em outgoing\/} edges. Thus, vector~$\Phi$ is identified with function
$$
\varphi=\varphi_a+\varphi_b.
$$
Now decompose this function as
$$
\varphi=\psi_a+\psi_b,
$$
where $\psi_a$ and $\psi_b$ belong to the spaces corresponding
to {\em incoming\/} edges, identify $\psi_a$ and $\psi_b$
with vectors $\Psi_a \in V_a$ and $\Psi_b \in V_b$ respectively
using the isomorphisms for incoming edges, and set
$$
\Psi=\Psi_a+\Psi_b.
$$
The operator $X_{ab}$ is defined by the requirement that
$$
X_{ab}\Phi=\Psi
$$
for every vector $\Phi$ (recall that $X_{ab}$ acts ``against the arrows'').

The equation (\ref{ieq2}) for $X$-operators constructed this way
follows from the fact that both LHS and RHS of~(\ref{ieq2}) act
on each ``outgoing'' vector~$\Phi$, where now
$\Phi\in V_a\oplus V_b\oplus V_c$,
as follows. First, $\Phi$ is identified with a meromorphic
function~$\varphi$ using isomorphisms corresponding to three outgoing
edges. Then $\varphi$ is identified with some vector
$\Psi\in V_a\oplus V_b\oplus V_c$
using three ``incoming'' isomorphisms. So, both LHS and RHS transform
$\Phi$ into $\Psi$, although using different ways of doing this.

The explicit formulae determining $X_{ab}$ are as follows. Recall that the
isomorphisms between the spaces~$V_a$ of column vectors and the spaces of
meromorphic functions are provided by fixing bases in the latter spaces.
Let $(f_1,\ldots,f_{n_a})$ and $(g_1,\ldots,g_{n_b})$ be bases corresponding
to {\em incoming\/} edges for some~$X_{ab}$, while $(\tilde f_1,\ldots,
\tilde f_{n_a})$ and $(\tilde g_1,\ldots,\tilde g_{n_b})$ be bases
corresponding to the {\em outgoing\/} edges. Then it is not hard to verify
that $X_{ab}$ is detrmined by the formula
$$
\left(f_1(z),\ldots,f_{n_a}(z),g_1(z),\ldots,g_{n_b}(z)
\vphantom{\tilde f}\right)X_{ab}=
\left(\tilde f_1(z),\ldots,\tilde f_{n_a}(z),\tilde g_1(z),\ldots,
\tilde g_{n_b}(z)\right)
$$
that must hold for all $z\in \Gamma$.

\medskip
{\bf Remark. }Our model, being composed of a finite number of strings
moving in the plane, has the obvious ``global'' integral of motion
(or rather all admissible motions)---the product of {\em all\/}
$X$-operators (complemented by unities in the lacking spaces)
taken in the order corresponding to moving along the strings
in positive directions. If, however, we want to ensure the possibility
of passing to the infinite string number limit, we should prefer
more ``local'' constructions. For example, using the presented method of
constructing algebro-geometric solutions, one can add
a new string---let us denote it $h$---so that
it don't intersect some chosen domain~$\Omega$ of the plane, adding
at the same time to divisor $\cal D$ one of the divisors ${\cal D}'_h$
or ${\cal D}''_h$, depending on whether $\Omega$ lies to the left or to the
right of~$h$. The result will be no change to the $X$-matrices
(as well as divisors corresponding to string segments) in~$\Omega$.
The reader can find some formulae for the infinite kagome lattice
in~\cite{korepanov dis}, chapter~3.
\medskip

\section{One-parametric solutions to the FTE}
\label{sec one-parametric}

The simplest case of the ``re-factorizing
equation''~(\ref{ieq pererazlozhenie}) is when the blocks in matrices
are just numbers. For numbers, we will use small letters, i.e.\ let
\begin{eqnarray}
X_{12}=
\left(\begin{array}{ccc}
a_1& b_1&  0\\
c_1& d_1&  0\\
0  & 0  &  1
\end{array}\right),&&
X'_{12}=
\left(\begin{array}{ccc}
a'_1& b'_1& 0\\
c'_1& d'_1& 0\\
   0&    0& 1
\end{array}\right),\nonumber\\
X_{13}=
\left(\begin{array}{ccc}
a_2& 0& b_2\\
  0& 1&   0\\
c_2& 0& d_2
\end{array}\right),&&
X'_{13}=
\left(\begin{array}{ccc}
a'_2& 0& b'_2\\
   0& 1&    0\\
c'_2& 0& d'_2
\end{array}\right),\nonumber\\
X_{23}=
\left(\begin{array}{ccc}
 1&   0&   0\\
 0& a_3& b_3\\
 0& c_3& d_3
\end{array}\right),&&
X'_{23}=
\left(\begin{array}{ccc}
1&    0&    0\\
0& a'_3& b'_3\\
0& c'_3& d'_3
\end{array}\right),\nonumber
\end{eqnarray}
and let us search for solutions of equation
\be
X_{12}X_{13}X_{23}=X'_{23}X'_{13}X'_{12}.
\label{s eq1}
\ee

The (nontrivial) {\em one-parametric\/} solutions of~(\ref{s eq1}) were
classified in~\cite{sergeev ferroelectro}. The word `one-parametric' means
here that $a_k$, $b_k$, $c_k$ and $d_k$, as well as
$a'_k$, $b'_k$, $c'_k$ and $d'_k$, are supposed to be algebraic
functions of {\em one\/} complex parameter~$x$ in such way that
\begin{eqnarray}
a_k=a(x_k),\quad b_k=b(x_k),\quad c_k=c(x_k),\quad d_k=d(x_k),&
\nonumber\\
a'_k=a(x'_k),\quad b'_k=b(x'_k),\quad c'_k=c(x'_k),\quad d'_k=d(x'_k),&&
k=1,2,3,
\nonumber
\end{eqnarray}
and the operator~$R$ can be regarded as a transformation
$$
R:\quad (x_1,x_2,x_3)\to (x'_1,x'_2,x'_3).
$$
It turns out that there are, essentially, six different cases for
the matrix
$$
X(x)=\pmatrix{a(x) & b(x)\cr c(x) & d(x)}.
$$

\subsection{Case $(\alpha)$}

$$
X(x)=\pmatrix{1 & x\cr 0 & k},
$$
$k$ being a constant; this gives
$$
R:\quad x_1,x_2,x_3 \rightarrow x_1,\;kx_2+x_1x_3,\;x_3.
$$
Inverse map:
$$
R^{-1}:\quad x_1,x_2,x_3\rightarrow
x_1,\; {x_2-x_1x_3\over k},\;x_3.
$$

\subsection{Case $(\beta)$}
\label{subs beta}

$$
X(x)=\pmatrix{1 & x\cr k/x & 0},
$$
this gives
$$
R:\quad x_1,x_2,x_3\rightarrow
{kx_2+x_1x_3\over x_3},\;x_1x_3,\;{kx_2x_3\over kx_2+x_1x_3}.
$$
Inverse map:
$$
R^{-1}:\quad x_1,x_2,x_3\rightarrow
{x_1x_2\over x_2+x_1x_3},\; {x_1x_3\over k},\;{x_2+x_1x_3\over x_1}.
$$

\subsection{Case $(\gamma)$}

$$
X(x)=\pmatrix{x & 0\cr 1-x & 1},
$$
then
$$
R:\quad x_1,x_2,x_3\rightarrow
{x_3-x_2+x_1x_2\over x_3},\;{x_1x_2x_3\over x_3-x_2+x_1x_2},\;x_3.
$$
Inverse map:
$$
R^{-1}:\quad x_1,x_2,x_3\rightarrow
{x_1x_2\over x_3-x_1x_3+x_1x_2},\;x_3-x_1x_3+x_1x_2,\;x_3.
$$

\subsection{Case $(\delta)$}

$$
X(x)=\pmatrix{x & 1\cr 1-x & 0}.
$$
Then
$$
R:\quad x_1,x_2,x_3\rightarrow
{x_1x_2\over x_1+x_3-x_1x_3},\;x_1+x_3-x_1x_3,\;
{(1-x_1)x_2x_3\over x_1+x_3-x_1x_2-x_1x_3}.
$$
Here
$$
R^2=1.
$$
This transformation is connected with the pentagon equation and
was described in \cite{tenterm}.

\subsection{Case $(\epsilon)$}
\label{subsec epsilon}

$$
X(x)=\pmatrix{x & 1+ix\cr 1-ix & x},
$$
then
\begin{equation}
R:\quad x_1,x_2,x_3\rightarrow
{x_1x_2\over x_1+x_3+x_1x_2x_3},\;
x_1+x_3+x_1x_2x_3,\;
{x_2x_3\over x_1+x_3+x_1x_2x_3},
\label{ieq electric transform}
\end{equation}
again with
$$
R^2=1.
$$
This is the electric network transformation, considered
in~\cite{kashaev electro}.
The variables $x_j$ and $x'_j$ are connected with the resistances
$r_j$ and $r'_j$ in the Introduction as follows:
$$
x_1={1\over r_1},\quad x_2=r_2,\quad x_3={1\over r_3},\quad
x'_1=r'_1,\quad x'_2={1\over r'_2},\quad x'_3=r'_3.
$$
The reader can verify that the equalness of results of two transformation
sequences in Figure~\ref{iris4} means exactly the same that the FTE
for $x$'s, if all the resistances attached to the edges of graphs
in Figure~\ref{iris4} are replaced by $x_j^{\pm 1}$, where the proper
choice of signs is an easy exercise.

Note also that in~\cite{kashaev electro} this transformation
was realized in terms of a ``free-bosonic local Yang--Baxter equation'',
while our case here is a fermionic one, as explained
in~\cite{korepanov dis,sergeev ferroelectro}.

\subsection{Case $(\zeta)$}

$$
X(x)=\pmatrix{x & -s(x)\cr s(x) & x},
$$
where $s(x)^2=1-x^2$.
This case is equivalent to Onsager star--triangle transform, and can
also be interpreted
as the Euler decomposition of an element of the group~$SO(3)$
into a product of two-dimensional rotations, if we put
$x=\cos\phi$, $s(x)=\sin\phi$.
This case was also considered in \cite{kashaev electro}.

For $R$, this gives
$$
R:\quad x_1,x_2,x_3\rightarrow
{x_1x_2\over F(x_1,x_2,x_3)},\;
F(x_1,x_2,x_3),\;{x_2x_3\over F(x_1,x_2,x_3)},
$$
where $F$ can be found from
$$
s(F)=s(x_2)x_1x_3-s(x_1)s(x_3),
$$
and so this is a two-foiled transformation.

Inverse map:
$$
R^{-1}=IRI,
$$
where $I$ transforms a pair $\bigl(x,s(x)\bigr)$ into
$\bigl(x,-s(x)\bigr)$.

\section{Electric network transformation and Miwa model}
\label{sec miwa}

It has been shown in paper~\cite{kashaev electro} that the ``electric''
model described in the Introduction and in subsection~\ref{subsec epsilon}
is connected with the well known integrable Miwa model~\cite{miwa}.
In terms of the present paper, this connection reads as follows.

Let us consider, in the three-dimensional space ${\bf R}^3\ni (x,y,z)$,
the cubic lattice planes whose equations are $x=l$, $y=m$, and $z=n$, where
$l,m,n$ take all integer values. Consider also a ``moving'' plane with
the equation
\be
x+y+z=t,
\label{ieq moving plane}
\ee
where $t$ can be thought of as ``time''. With a generic~$t$, the
intersection of the cubic lattice planes with the
plane~(\ref{ieq moving plane}) yields in this latter the regular
(infinite) kagome lattice. Let us choose some positive direction
for every straight line forming that lattice in such manner that all
parallel lines be directed the same way and the orientation of all the
lines be consistent in the sense of section~\ref{sec straight-string}.

Let us now attach to each kagome lattice vertex a variable~$x_k$, and
require that the strings move with time~$t$ according to
equation~(\ref{ieq moving plane}) and the variables~$x_k$ change
according to~(\ref{ieq electric transform}) whenever a vertex passes
through a line. In such way, we will obtain a straight-string model
of the kind described in section~\ref{sec straight-string}.

Let us introduce unit
vectors $f_1,f_2,f_3$ pointing in the directions of axes $x,y,z$ of
the space~${\bf R}^3$ respectively.
Note that each kagome lattice vertex sweeps
between its two collisions with strings
exactly an edge of the integer cubic lattice in~${\bf R}^3$.
Thus, we can consider values~$x_k$ as corresponding to edges of the
cubic lattice. For convenience, and also in order to link our
presentation
to the paper~\cite{kashaev electro}, we will think of $x_k$'s also
as attached to {\em vertices\/} of the cubic lattice as follows:
let the radius vector
of some vertex be~$n$, then we will denote as $x_1(n)$ the value~$x_1$
corresponding to the edge linking the vertices with
radius vectors $n$ and $n-f_1$, as $x_2(n)$---the value~$x_2$
corresponding to the edge linking the vertices
$n+f_2$ and $n$, and  as $x_3(n)$---the value~$x_3$
corresponding to the edge linking the vertices $n$ and $n-f_3$.

Let us attach signs $\epsilon(n)=\pm 1$ to the cubic lattice vertices
in a checkerboard order. Then let us
introduce (again in order to link our paper to the
work~\cite{kashaev electro}) the vectors $e_1,e_2,e_3$
according to
\be
e_1={f_3-f_2-f_1\over 2},\quad e_2={f_1+f_2+f_3\over 2},\quad
e_3={f_1-f_2-f_3\over 2}.
\label{ieq e-f}
\ee
Finally, fix some non-zero constants $\alpha_1,\alpha_2$ and $\alpha_3$
and make a substitution
$$
x_j(n)=\epsilon(n)\,\alpha_j\,
{\tau(n+e_1+e_2+e_3)\,\tau(n-e_j)\over
\tau(n+e_k)\,\tau(n+e_l)},
\qquad \{j,k,l\}=\{1,2,3\},
$$
where $\tau$ is the new unknown function, defined, as is seen from
(\ref{ieq e-f}), in the {\em centers of cubes\/} of the lattice.

As is shown in \cite{kashaev electro}, function~$\tau$
satisfies the {\em Miwa equation\/}
$$
\sum_{j=1}^4 \alpha_j \tau(n+e_j) \tau(n-e_j)=0,
\qquad e_4=-e_1-e_2-e_3,
\qquad \alpha_4=\alpha_1 \alpha_2 \alpha_3.
$$
Note that the variables $a_j(n)$ in \cite{kashaev electro} are connected
with our $x_j(n)$ by
$$
a_j(n)=\epsilon(n) x_j(n)^{-1}.
$$

\section{Quantization of the $(\beta)$ model}
\label{sec quant}

The model corresponding to the case~$(\beta)$ (subsection~\ref{subs beta})
admits an amazingly straightforward quantization. According to
subsection~\ref{subs beta}, the corresponding transformation
$(x_1,x_2,x_3)\to (x'_1,x'_2,x'_3)$ arises from the relation
\begin{eqnarray}
\pmatrix{1 & x_1 & 0\cr k/x_1 & 0 & 0\cr 0 & 0 & 1}
\pmatrix{1 & 0 & x_2\cr 0 & 1 & 0\cr k/x_2 & 0 & 0}
\pmatrix{1 & 0 & 0\cr 0 & 1 & x_3\cr 0 & k/x_3 & 0}
\nonumber\\[\smallskipamount]
=\pmatrix{1 & 0 & 0\cr 0 & 1 & x'_3\cr 0 & k/x'_3 & 0}
\pmatrix{1 & 0 & x'_2\cr 0 & 1 & 0\cr k/x'_2 & 0 & 0}
\pmatrix{1 & x'_1 & 0\cr k/x'_1 & 0 & 0\cr 0 & 0 & 1}.
\label{ieq matr}
\end{eqnarray}

To quantize this model, let us take the same relation~(\ref{ieq matr}),
but with $x_1,x_2$ and $x_3$ belonging to some associative algebra
and satisfying the commutation relations
\be
x_1 x_2=\omega^{-1}x_2 x_1,\qquad
x_1 x_3=\omega\, x_3 x_1,\qquad
x_2 x_3=\omega^{-1}x_3 x_2,
\label{ieq comm}
\ee
where $\omega$ is a scalar (and $k$ remains a scalar, too). It can be
verified that in this case (\ref{ieq matr}) still determines the ``primed''
variables without contradiction, namely
\be
x'_1=x_1+kx_2x_3^{-1},\qquad x'_2=x_1x_3,\qquad
x'_3=k(x_1+kx_2x_3^{-1})^{-1}x_2,
\label{ieq 'noncomm}
\ee
and the following relations hold:
\be
x'_1 x'_2=\omega\, x'_2 x'_1,\qquad
x'_1 x'_3=\omega^{-1} x'_3 x'_1,\qquad
x'_2 x'_3=\omega\, x'_3 x'_2.
\label{ieq comm'}
\ee

Let us consider now a sraight-string model, as in
section~\ref{sec straight-string}, and define the commutation relations
for $x_k$ in {\em all\/} vertices. We will demand that
\begin{itemize}
\item
if two $x_k$ don't lie in the same string, they commute;
\item
if the vertices where two $x_k$ are situated belong to a triangle
of the form as in Figure~\ref{ifig quant} (a) or~(b),
\begin{figure}[htp]
\begin{center}
\unitlength=0.80mm
\special{em:linewidth 0.4pt}
\linethickness{0.4pt}
\begin{picture}(128.00,71.00)
\put(0.00,43.00){\line(2,-1){50.00}}
\put(5.00,23.00){\line(3,2){54.00}}
\put(35.00,8.00){\line(1,3){21.00}}
\put(83.00,8.00){\line(1,3){21.00}}
\put(73.00,13.00){\line(3,2){54.00}}
\put(78.00,63.00){\line(2,-1){50.00}}
\put(98.00,53.00){\circle*{1.25}}
\put(118.00,43.00){\circle*{1.25}}
\put(88.00,23.00){\circle*{1.25}}
\put(40.00,23.00){\circle*{1.25}}
\put(50.00,53.00){\circle*{1.25}}
\put(20.00,33.00){\circle*{1.25}}
\put(4.00,41.00){\vector(2,-1){2.00}}
\put(36.00,11.00){\vector(1,3){1.00}}
\put(84.00,60.00){\vector(2,-1){2.00}}
\put(84.00,11.00){\vector(1,3){1.00}}
\put(20.00,36.00){\makebox(0,0)[cb]{$1$}}
\put(44.00,25.00){\makebox(0,0)[lc]{$2$}}
\put(51.00,51.00){\makebox(0,0)[lt]{$3$}}
\put(94.00,51.00){\makebox(0,0)[rc]{$2$}}
\put(117.00,46.00){\makebox(0,0)[cb]{$3$}}
\put(90.00,21.00){\makebox(0,0)[lt]{$1$}}
\put(8.00,25.00){\vector(3,2){3.00}}
\put(76.00,15.00){\vector(3,2){3.00}}
\put(28.00,1.50){\makebox(0,0)[cc]{(a)}}
\put(100.00,1.50){\makebox(0,0)[cc]{(b)}}
\end{picture}
\end{center}
\caption{}
\label{ifig quant}
\end{figure}
the commutation relations~(\ref{ieq comm}) hold (where we temporarily
change the actual $k$'s to 1, 2 and 3). Note that there may
exist other strings, not depicted in Figure~\ref{ifig quant},
between the vertices 1, 2 and 3;
\item
if the vertices where two $x_k$ are situated arise as intersection points
of a string with two parallel strings, the commutation relation between
them can be obtained if we let one of the vertices in Figure~\ref{ifig quant}
tend to infinity. Here the figures (a) and~(b) lead to the same result,
however the orientation of strings does play the r\^ole.
\end{itemize}

It can be verified that the stated commutation relations are conserved
under the transformations as in Figure~\ref{ifig1}.

\section{Discussion}
\label{discussion}

In this paper, we have considered, first of all, classical integrable
dynamical systems in $2+1$-dimensional discrete space-time. We proposed
a very general scheme for constructing such systems together with their
solutions. We showed that our scheme includes Miwa bilinear equation
as just one of particular cases.

The integrability in our general scheme is ensured by a rich set of
symmetries that are expressed in the local form as functional tetrahedron
equation (FTE). The form of FTE suggests at once that the corresponding
model need not be associated with a regular (e.g.\ cubic) lattice, but
can be naturally defined e.g.\ for any configuration of intersecting planes
(``world sheets'' of strings in section~\ref{sec straight-string})
in a three-dimensional space. The algebro-geometric solution for such system
is given in terms of an algebraic curve and some constant divisors assigned
to the planes. In a sense, the equations of motion, when expressed in terms
of those divisors, ``dissappear'', and this calls to mind the Penrose
twistor theory, where field equations ``dissappear'' as well.

We have also shown that, at least in one particular case, FTE can
successfully replace the usual (quantum) tetrahedron equation in
constructing a quantum integrable model.

Finally, we would like to note that a generalization onto $3+1$
dimensions of our scheme for
constructing classical models and their algebro-geometric solutions
has been found recently in work~\cite{korepanov 3+1}.

\end{document}